\documentclass[printer]{aa}
\usepackage{amsmath,amssymb}
\usepackage{subfigure}        
\usepackage{txfonts}
\usepackage{multirow}
\usepackage{float}
\usepackage{longtable}
\usepackage{graphicx}
\usepackage{hyperref}
\usepackage{threeparttable}

\begin{document}

\title{Measuring three-dimensional shapes of stable solar prominences using stereoscopic observations from SDO and STEREO}
\author{Chengrui Zhou\inst{\ref{inst1},\ref{inst2}} \and Chun Xia\inst{\ref{inst1},\ref{inst3}} \and Yuandeng Shen\inst{\ref{inst2}}}
\institute{School of Physics and Astronomy, Yunnan University, Kunming 650500, China \email{chun.xia@ynu.edu.cn}\label{inst1} 
           \and Yunnan observatories,Chinese Academy of sciences, Kunming 650216, China\label{inst2}
           \and National Astronomical Observatories, Chinese Academy of Sciences, Beijing 100101, China\label{inst3}}

\abstract{}
{Although the real shapes and trajectories of erupting solar prominences in three dimensions 
 have been intensively studied, the three-dimensional (3D) shapes of stable prominences before eruptions 
have not been measured accurately so far. We intend to make such a measurement to constrain 3D prominence models 
and to extend our knowledge of prominences.}
{Using multiperspective observations from the Atmospheric Imaging Assembly on board the Solar Dynamics Observatory (SDO) and
the Extreme Ultraviolet Imager on board the Solar Terrestrial Relations Observatory (STEREO),
we reconstructed 3D coordinates of three stable prominences: a quiescent, 
an intermediate, and a mixed type. Based on the 3D coordinates, we measured the height, length, and 
inclination angle of the legs of these prominences.
To study the spatial relationship between the footpoints of prominences and photospheric magnetic 
structures, we also used the Global Oscillation Network Group H$\alpha$ images and magnetograms from the Helioseismic and Magnetic Imager on board the SDO.}
{In three stable prominences, we find that the axes of the prominence legs are inclined by 68$\pm$6 degrees on average to the solar surface. Legs at different locations 
along a prominence axis have different heights with a two- to threefold difference.
Our investigation suggests that over 96\% of prominence footpoints in a sample of 70 footpoints are located at supergranular boundaries. The widths of two legs have similar values 
measured in two orthogonal lines of sight. We also find that a prominence leg 
above the solar limb showed horizontal oscillations with larger amplitudes at higher locations.}
{With a limited image resolution and number of cases, our measurement suggests that the legs of 
prominences may have various orientations and do not always stand vertically on the surface of the sun. Moreover, the locations of prominence legs are closely related to supergranules.}
{}


\titlerunning{Measuring three-dimensional shapes of stable solar prominences}
\authorrunning{C.R. Zhou et al}

\maketitle

\section{Introduction}\label{intro}
Solar prominences, also called solar filaments, consist of relatively cool 
and dense plasma \citep[see, e.g.,][]{Labrosse2010,Parenti2014}. They are supported by helical
magnetic flux ropes or sheared magnetic loops in the tenuous and hot solar corona 
above photospheric magnetic neutral lines \citep[see, e.g.,][]{Mackay2010}.
Prominences in quiescent regions, known as quiescent prominences, often consist 
of several discrete pillar-like structures~\citep{Panesar2014} that are variously referred to as ``legs"~\citep{Wedemeyer2013}, ``feet"
~\citep{Aulanier1998a,Shen2015}, ``tornadoes"~\citep{Pettit1943,Su2012,Panasenco2014}, and 
``barbs"~\citep{Martin1998}. These pillar-like structures are called legs in 
this paper. Prominence legs are often linked by a long continuous body, the so-called spine,
in the upper part in a well-developed prominence, and the lower parts of the prominence legs 
become narrow and sharp when they approach the chromosphere~\citep{Martin1998}. 
Sometimes, especially in high-latitude regions or close to the solar limb when the line-of-sight (LOS) 
integration path through the spine is short, the spine of a prominence is almost 
invisible in the H$\alpha$ line, and only prominence legs remain~\citep{Su2012,Li2013}. 
The formation of filaments begins with a few individual legs, which later combine to form
a continuous body \citep{Pevtsov2005}. Prominence legs appear to be fundamental building blocks
 of prominences.

The location of prominence legs is likely related to supergranulation, which can be traced by 
the chromospheric network and the photospheric magnetic network \citep{Simon1964}. 
By comparing H$\alpha$ prominence legs with Ca II K chromospheric network, 
\citet{Plocieniak1973} found that 90\% of the footpoints of 
filament legs are located at boundaries of supergranular cells, which was confirmed by
\citet{Pevtsov2005} using off-band H$\alpha$ chromospheric network images. \citet{Lin2005} derived 
photospheric flow cells representing supergranular cells and found that about 65\% of the
endpoints of filament barbs are located at supergranular cell boundaries. Morever, the endpoints
of filament barbs may not be the footpoints of legs, as shown by \citet{Su2012}
in their Figure 3 from two viewing angles at the same time, where some legs are present in side views
but are hidden in top-down views, and barbs shown in top-down views are not visible in side views. The
one-to-one correspondence between legs and barbs has been doubted, and some dynamic barbs may be due to 
longitudinal oscillations of filament threads high above the solar surfaces \citep{Ouyang2020}. Without
3D information, it is difficult to accurately identify the footpoints of the legs.
The relation between supergranulation and prominence legs needs further investigation, which may shed
light on the unclear magnetic structure of prominence legs.

Because the gravitational scale height of prominence plasma is shorter than the height of prominences, which means the
gas pressure gradient cannot be the main support for prominence plasma, it is commonly accepted that
prominence plasma is supported by the Lorentz force in magnetic dips \citep{Mackay2010}. Three-dimensional static force-free 
magnetic field models have been developed to interpret a filament barb as magnetic dips piled up from
nearby parasitic magnetic elements on the solar surface to the main magnetic flux rope along the filament axis 
\citep{Aulanier1998a,Aulanier1998b}. This interpretation of a leg as assembled lateral dips was 
supported by a high-resolution observation from one viewing angle of an active region filament, 
whose barbs terminated above the small polarity inversion line dividing two close magnetic elements of 
opposite polarities \citep{Chae2005}. \citet{Gunar2018} filled the magnetic dips of thousands of magnetic 
field lines from a 3D prominence force-free magnetic field model with hydrostatic prominence plasma \citep{Gunar2013} 
and constructed radiative-transfer-based synthetic H$\alpha$ images from a filament and prominence viewing angles.
The authors found that the collected magnetic field lines showing magnetic dips up to a pressure scale height are similar to
the synthetic H$\alpha$ images in the filament view, but occupy a wider volume than the synthetic H$\alpha$ structures
in the prominence view. To verify and constrain realistic 3D prominence models, it is important to accurately 
measure the 3D shapes of prominences, including the legs. However, measurements like this have large uncertainties from projection effects and can only be improved by 
time-series images because the solar rotation changes the relative viewing angle of 
prominences~\citep{Rosa1996} before 2006 when ground- and space-based 
telescopes were only able to observe from a single solar-terrestrial angle of view. 
Since 2006, when the Solar TErrestrial RElations Observatory (STEREO) Ahead (A) and Behind (B) spacecraft
~\citep{Kaiser2008} were launched, the two satellites, allowing a stereoscopic vision
of the Sun from multiple viewing angles, have started uncovering instantaneous 3D structures
of solar activities through triangulation and reconstruction methods
~\citep{Inhester2006,Feng2007,Aschwanden2008,Thompson2009}. Using these stereoscopy 
techniques, 3D information of an erupting active region filament on May 19, 2007, 
were studied~\citep{Gissot2008,Liewer2009,Xu2010},  
when the separation angle between STEREO A and B was small, for instance, 8.5 degrees. 
Simultaneous heating of the rising filament and the chromosphere below
were identified~\citep{Liewer2009}. A four-hour slow-rising phase with an upward 
acceleration of the filament was 
found before its impulsive eruption~\citep{Xu2010}. \citet{Bemporad2009}
estimated the expansion factor in 3D of an erupting active 
region prominence and found that the early expansion is anisotropic and mainly in the
radial direction, with an overall nonrotating ribbon-like shape. 
\citet{Liu2012} investigated the partial 
eruption of a double-decker active region filament with 3D reconstruction and 
inferred that the upper and lower branch of the filament had negative helicity. 
\citet{Zhou2017} tracked 3D structures from gradual to the impulsive phase of 
an erupting active region prominence and found a shape transformation of the prominence 
from a sigmoid shape to a loop arcade. All these studies are focused on 3D shapes of erupting 
active region filaments. The 3D shapes of nonerupting stable prominences, especially 
those in quiescent regions, have not been measured accurately, as far as we know. 

Based on stereoscopic observations from two viewing angles, 
we therefore report our measurement of 3D shapes of three stable prominences with a reconstruction of the 3D coordinates, and we further investigate the spatial relation between the footpoints of 
prominence legs and supergranular boundaries.
After an introduction in Section \ref{sec:data} of the instruments and data we used, we
describe the 3D reconstruction technique in Section \ref{sec:tec}. The results for 
the three prominences we selected are shown in Section \ref{sec:res}. Conclusions are 
summarized and discussed in Section \ref{sec:con}.

\section{Instruments and data}\label{sec:data}

To explore the effects of solar activity on the Earth, NASA launched
a series of solar space telescopes, such as the STEREO and the
Solar Dynamics Observatory (SDO) \citep{Pesnell2012}. 
The Atmospheric Imaging Assembly (AIA) \citep{Lemen2012} on board the SDO provides 
full-disk images out to 1.5 solar radii with 1.5$\arcsec$ spatial resolution 
(0.6$\arcsec$ pixel size) and 12-second temporal resolution in seven extreme 
ultraviolet (EUV) and two ultraviolet wavebands and in one visible light waveband.
We used AIA EUV images in 304 \AA~(0.05 MK),  171 \AA~(0.63 MK), and 193 \AA~
(1.6 MK and 20 MK) wavelength bands.
The Extreme Ultraviolet Imager (EUVI)~\citep{Wuelser2004} of the
SECCHI instrument suite \citep{Howard2008} on board the two STEREO
spacecraft observes the Sun from two viewing angles in four EUV spectral channels 
with 1.6$\arcsec$ pixel size and a full-Sun field of view out to 1.7 solar radii. 
We used 171 \AA~(0.63 MK) EUVI images. We also used full-disk LOS photospheric 
magnetograms provided by the Helioseismic and Magnetic Imager (HMI) \citep{Schou2012} 
on board the SDO with a 0.5$\arcsec$ pixel size and a cadence of 720 s.

To search suitable data for 3D reconstruction, we focused on the time at which prominences
can be observed simultaneously from different angles by at least two spacecraft, 
including the SDO and  STEREO A or B, and the angle between the lines of sight of 
the two spacecraft should be less than 130 degrees. We examined all data from 
October 2010 to March 2013, mainly looking at dark features of filaments in
171\AA~images of SDO/AIA and STEREO/EUVI, and in H$\alpha$ 
full-disk images from the Global Oscillation Network Group (GONG). Because the available images have a limited resolution, we targeted relatively large stable prominences 
with clear structures. In the end, we selected three prominences that we call 
PA, a quiescent prominence on February 9, 2012; PB, which is an intermediate prominence on May 25, 
2012; and PC, a quiescent-intermediate mixed prominence on November 19, 2011. PA
was observed by STEREO A and SDO separated by 115.6 degrees in viewing angles. PB was observed by
STEREO B and SDO separated by 115.85 degrees in viewing angles. PC was observed by STEREO A and
SDO separated by 106.1 degrees in viewing angles.

\section{3D reconstruction and data analysis}\label{sec:tec}

When the apparent positions of an object are determined 
in two images observed by two known observers from different 
viewing angles, then the 3D coordinates of the object can be derived with the 
tie-pointing method ~\citep{Thompson2006,Thompson2009}. A 3D reconstruction
 program using the method is ready to use in the SolarSoftware (SSW)~\citep{Freeland1998}. 
We used {\em scc\_measure.pro} in SSW to manually obtain 
the 3D coordinates and image pixel positions of features of
the prominences. The specific steps are as follows. First, we opened the two 171 \AA~images, 
called A and B, from AIA or EUVI. Second, we selected a feature of the targeted structure
in image A. The LOS of image A that passes through this point then shows up as a straight 
line in image B. Third, we selected a point that was most likely the same structure in the line 
in image B. In the end, the program calculates and outputs the solar Carrington heliographic 
coordinates of the structure and its pixel positions in the two images.
When the LOS of image A passes through a thick filament structure in image B, then 
it is hard to determine the corresponding point in the filament segment on the line.
Filament structures closer to the underlying chromosphere usually have narrower extension 
and therefore smaller attached uncertainty in a 3D reconstruction. We therefore marked all 
distinguishable filament footpoints. However, dark structures of the
chromosphere near filament footpoints sometimes obscured the identification. 
Because low filament footpoints in images A and B are expected to have 
nearly the same Carrington coordinates, we plot Carrington 
coordinates latitude and longitude lines through the footpoints. 
To plot the same Carrington coordinate 
latitude and longitude lines on two images from different viewing angles,
we used {\em hel2arcmin.pro} in SSW to calculate the angular distance
from the given Carrington coordinates to the center of a given image.
We used one pixel size of a STEREO/EUVI image, that is, 870 km, as the uncertainty to
estimate the error of the 3D reconstruction, which entered all quantities derived 
from 3D reconstructed coordinates.

\section{Results}\label{sec:res}
\subsection{Quiescent prominence: PA}\label{sec:pa}
\begin{figure*}
 \includegraphics[width=\textwidth]{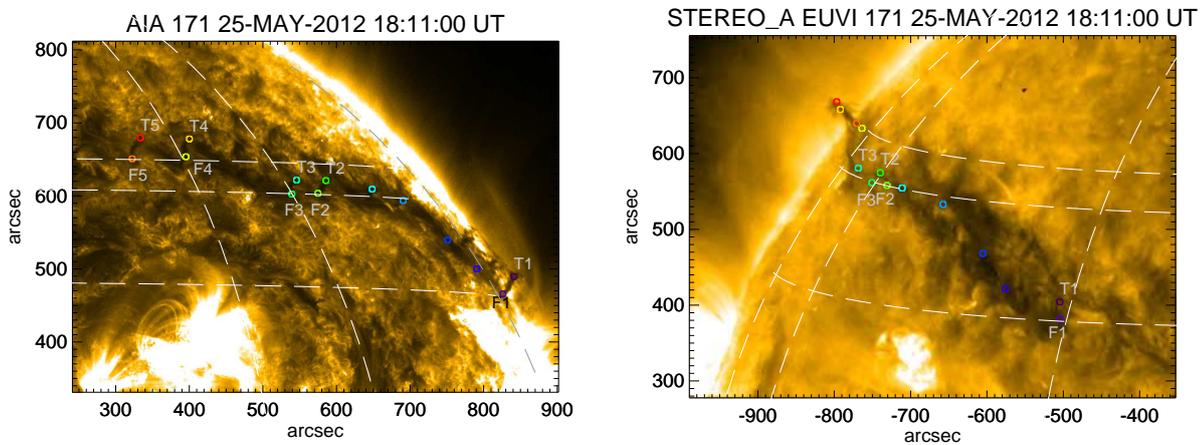}
 \caption{Prominence PA in the SDO/AIA 171\AA~image (left) and STEREO A/EUVI 171 
 \AA~image (right) at the same time on May 25, 2012. The EUVI image is in log-scale
for clarity. Different features are marked with circles in different colors, 
and circles of the same color in the two images represent the same structures.
The diameter of these circles is 8 arcsec (about 5,800 km).
 The dotted lines are the Carrington coordinate latitude and
 longitude lines, using the latitudes and longitudes of the three points F1, 
 F2, and F3. The LOS of STEREO A in the left picture is along the latitude, and 
 the dashed lines and arrows in the right image show the LOS of SDO.
}
 \label{fig:pa}
\end{figure*}

Prominence PA was observed on May 25, 2012, when the STEREO A
and SDO had a difference in viewing angles of 115.6 degrees.
As shown in Figure~\ref{fig:pa}, the prominence was close to the west
solar limb in the SDO/AIA 171 \AA~image, while it appeared at the east solar
limb in the STEREO/EUVI 171 \AA~image, which is shown in log scale to present 
the details of dark filament structures. The prominence consisted of more than
seven dark legs distributed along the magnetic neutral line from northeast 
to southwest, and a sparse filament spine linked these legs. The legs
are easier to distinguish in the SDO view by watching the flank of the prominence
than in the STEREO A view, which looked down at the prominence along its axis.
Based on the method described in section~\ref{sec:tec}, we determined 
the 3D coordinates of several features of the prominence. 
These features are marked with small circles in different 
colors. Small circles with the same color in the two images correspond to 
the same prominence structure. Five distinguishable legs of the 
prominence are represented by their footpoints and top points. 
The footpoints are labeled F1, F2, F3, F4, and F5, and the corresponding 
top points are labeled T1, T2, T3, T4, and T5. The legs are labeled 
F1-T1, F2-F2, and so on. In the STEREO A image, the legs F4-T4 
and F3-T3 almost overlap, so they are not labeled to avoid confusion. 
Four blue circles between F1 and F2 represent the bottom parts of the 
filament segment where the corresponding top parts are hard to identify in 
the EUVI image because of the low resolution and the disturbance from nearby dark 
chromospheric patches.

Using information from the 3D reconstruction, we can evaluate the true spatial 
sizes of the prominence without uncertainties from the projection effect.
In Table~\ref{table:1} we list the length of the prominence legs, 
the inclination angle of the legs, the height of the top points of the 
legs, and the height of the footpoints of the legs for all three prominences. 
The length of each leg is evaluated as the 
distance between the footpoint and the top point of the leg. The 
inclination angle of each leg is calculated as the angle between the 
foot-to-top line and the solar surface. When a leg is vertical to the solar 
surface, its inclination angle is 90 degrees. The average inclination 
angle of the five legs of PA is about 69 degrees. The least inclined leg 
F4-T4 has an inclination angle of 82$\pm$8 degrees. The most inclined
leg F1-T1 has a inclination angle of 54$\pm$12 degrees.
 The axes of legs F2-T2, F3-T3, and F4-T4 are
not straight, with an apparent bending near footpoints in the AIA image. 
These legs did not stand perpendicular to the solar surface, for instance,
leg F4-T4 is inclined to the north and leg F3-T3 is inclined to the
east. The highest point T3 of the PA has a height of 30,964$\pm$745 km. By 
accumulating the distances between the footpoints, we find that the 
overall length of the prominence is about 558,000 km.

\begin{table*}[htbp]
 \centering
 \begin{threeparttable}
 \caption{Geometrical properties of prominence legs obtained by 3D reconstruction.}\label{table:1}
   \begin{tabular}{clclll}
    \hline\hline   
    prom. legs & length (km) & Inclination ($^{\circ}$)\tnote{1} & T height (km) & F height (km) & F projection (km)\tnote{2} \\
   \hline
     PA F1-T1  &   21,438$\pm$4,430   &   54$\pm$12  &   22,852$\pm$111 &   5,375$\pm$132    & 5,281\\
     PA F2-T2  &   15,146$\pm$586     &   67$\pm$05  &   15,812$\pm$647 &   1,852$\pm$487    & 1,611\\
     PA F3-T3  &   18,257$\pm$459     &   70$\pm$04  &   17,310$\pm$633 &     118$\pm$557    & 100\\
     PA F4-T4  &   25,813$\pm$493     &   73$\pm$03  &   30,964$\pm$745 &   5,828$\pm$585    & 4,602\\
     PA F5-T5  &   27,122$\pm$788     &   82$\pm$08  &   30,637$\pm$842 &   3,746$\pm$786    & 2,812\\
     \\
     PB F1-T1  &   22,435$\pm$3,035    &   52$\pm$08  &   21,014$\pm$647 &   3,391$\pm$690   & 2,439\\
     PB F2-T2  &   21,330$\pm$683      &   86$\pm$09  &   21,331$\pm$800 &   1,825$\pm$960   & 1,233\\
     PB F3-T3  &   62,581$\pm$1,831    &   65$\pm$04  &   60,181$\pm$1,051&  3,330$\pm$587   & 1,883\\
     PB F4-T3  &   62,887$\pm$2,203    &   60$\pm$04  &   60,181$\pm$1,051&  5,270$\pm$646   & 3,083\\
     \\
     PC F1-T1  &   32,191$\pm$1,593    &   61$\pm$06  &   37,607$\pm$738 &   9,409$\pm$801   & 6,391\\
     PC F2-T2  &   26,590$\pm$1,518    &   63$\pm$08  &   31,250$\pm$724 &   7,513$\pm$1,155 & 5,156\\
     PC F3-T3  &   25,710$\pm$1,203    &   76$\pm$08  &   29,913$\pm$954 &   4,950$\pm$863   & 3,428\\
     PC F4-T4  &   24,252$\pm$152      &   81$\pm$03  &   28,395$\pm$835 &   4,400$\pm$696   & 3,076\\
     PC F5-T5  &   23,753$\pm$2,339    &   61$\pm$08  &   21,320$\pm$800 &     473$\pm$835   & 340\\
     PC F6-T6  &   50,203$\pm$1,905    &   68$\pm$04  &   45,468$\pm$821 &   1,065$\pm$814   & 813\\
     PC F7-T7  &   41,055$\pm$1,476    &   26$\pm$02  &   34,237$\pm$369 &  15,652$\pm$473   & -\\
   \hline
  \end{tabular}
  \begin{tablenotes}
   \item[1] The angle between a leg axis and the solar surface in units of degree.
   \item[2] The projected distance on HMI magnetograms of a prominence footpoint.
  \end{tablenotes}
 \end{threeparttable}
\end{table*}

\begin{figure*}
 \includegraphics[width=\textwidth]{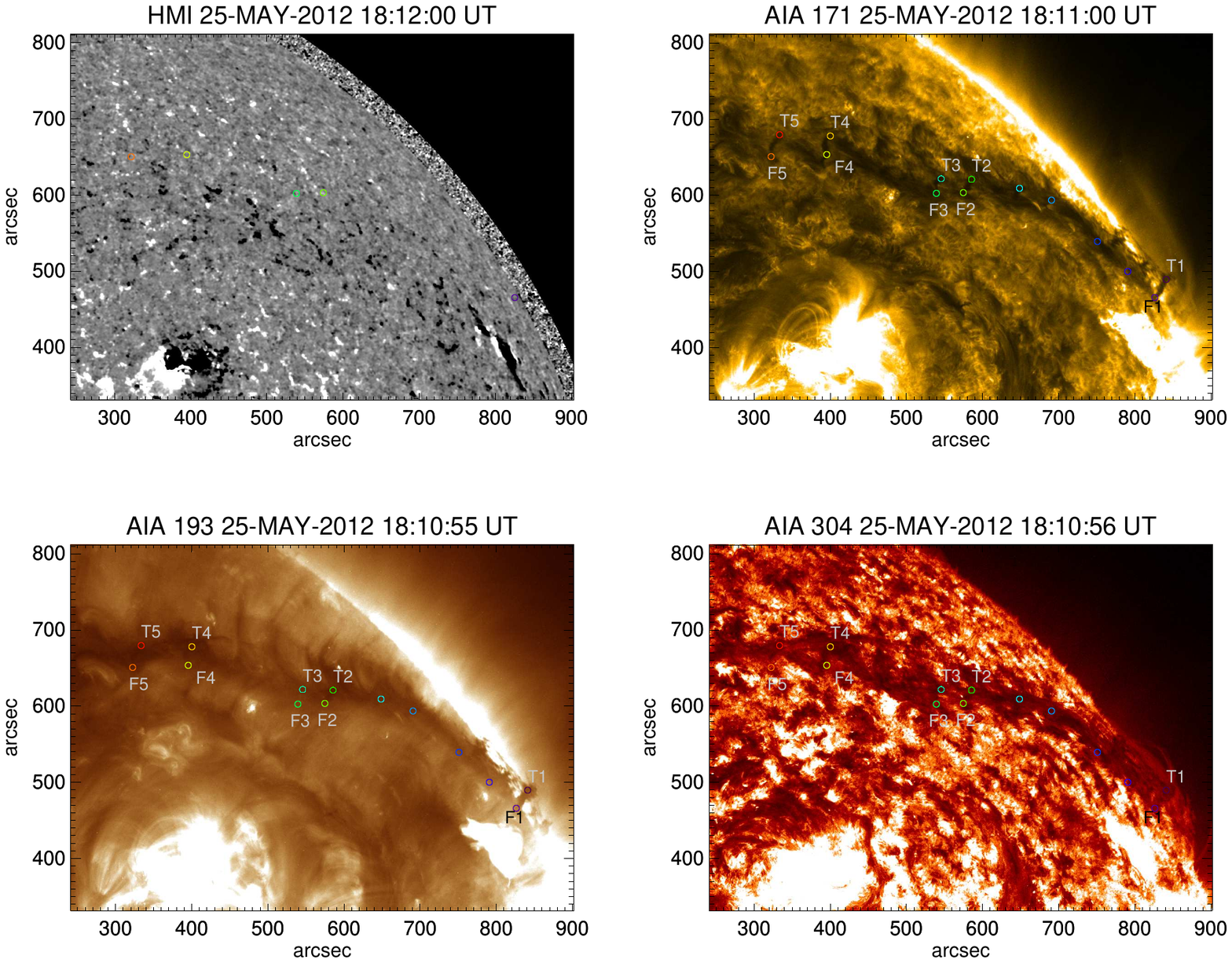}
 \caption{HMI magnetogram, AIA 171 \AA, 193 \AA, and 304 \AA~ images with the same
field of view looking at prominence PA on May 25, 2012. Feature points of the 
prominence are marked with circles in different colors for all images. 
The diameter of these circles is 8 arcsec (about 5,800 km).
The magnetogram is shown in gray saturated at $\pm45$ G.}
 \label{fig:pam}
\end{figure*}

To understand the magnetic field environment of the prominence 
legs, we compared the HMI photospheric magnetogram around PA with the
AIA 171 \AA, 193 \AA, and 304 \AA~images in the same field of view, as shown in 
Figure~\ref{fig:pam}. The magnetogram is shown with saturation values of $\pm$45 G.
The four reconstructed footpoints F2, F3, F4, and F5 seem to be located near polarity-mixed 
magnetic flux elements on supergranular boundaries. F1 is almost on the solar limb of the 
magnetogram.  The errors in magnetic field and projected position are too large.

The dark filament structures in the 171 \AA~and 193 \AA~EUV channels are mainly caused 
by absorption of background EUV emission
by neutral hydrogen, neutral helium, and singly ionized helium of filament plasma through
photoionization, with a minor contribution from the volume- blocking effect for on-disk 
features \citep{Anzer2005}. Compared to the 171 \AA~image, the legs on 
the solar disk are hardly visible in the 193 \AA~image, with very poor
contrast because of the strong foreground emission that is caused by the large coronal emission scale height \citep{Parenti2012}.
However, the absorption features above the solar limb around
leg F1-T1 is very similar in the 171 \AA~and 193 \AA~images. The bright structure on the 
top of leg F1-T1 in the 171 \AA~channel is due to the emission of Fe IX 171.07 \AA~line from the 
prominence-corona transition region (PCTR) around 400,000 K \citep{Parenti2012,DelZanna2011}.
In the 304 \AA~image, the dark legs on the disk are vaguely visible due to dark chromospheric features nearby. 
The 304 \AA~channel is dominated by the He II 303.78 \AA~line 
from 50,000 K plasma \citep{Odwyer2010}, which is the PCTR
covering prominence plasma. Cold prominence plasma shown as dark legs above solar limb in the 171 \AA~and 193\AA~channels is
hidden behind the emission from the large PCTR region in the 304 \AA~channel.
Above T1, the top of the 171 \AA~dark prominence, there are additional 
304 \AA~bright prominence structures. \citet{Wang1998} found a similar phenomenon with
 additional He II 303.78 \AA~emissions above H$\alpha$ emissions and 195 \AA~absorptions 
in a quiescent prominence. Because of the high optical thickness of the He II 303.78 \AA~line, the 304 \AA~structures are located in a layer of PCTR in front of the prominence with a spatially
extended shape \citep{Gunar2014}, where the low optical thickness H$\alpha$ line may render the contrast too high to be visible because the difference in integral depth along the LOS is too great.

\begin{figure*}
\includegraphics[width=\textwidth]{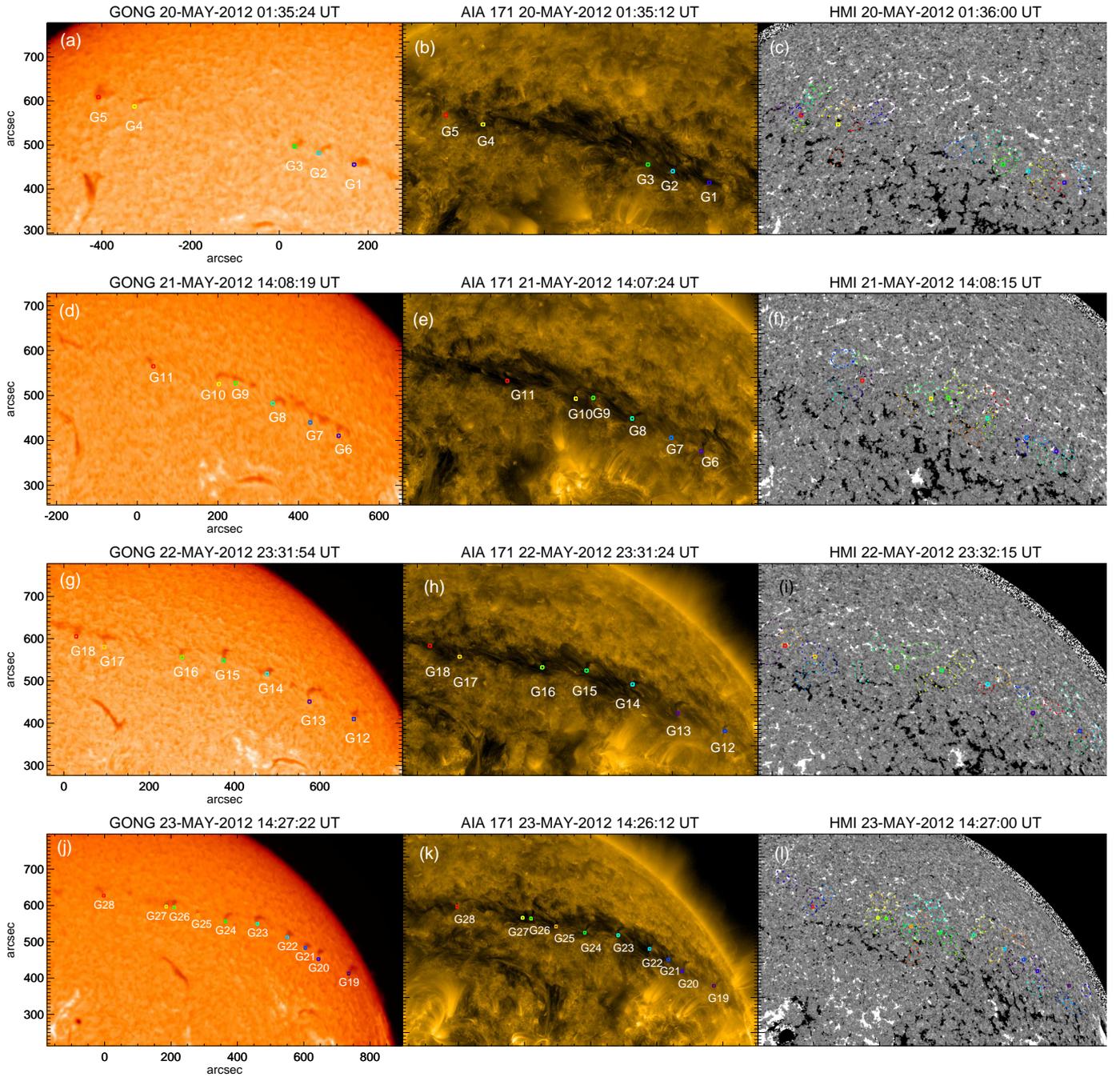}
\caption{Time sequence of prominence PA in GONG H$\alpha$ images, AIA 171 \AA~images in log scale, 
and Gaussian smoothed HMI LOS magnetograms on May 20 ((a)-(c)), May 21 ((d)-(f)), May 22 ((g)-(i)), and May 23 ((j)-(l))
in 2012. The small circles in different colors show the location of the footpoints of the prominence legs.
The diameter of these circles is 8 arcsec (about 5,800 km).
The magnetograms are shown in gray saturated at $\pm30$ G. Supergranular boundaries are indicated by
dashed lines in different colors connecting strong magnetic elements in the HMI magnetograms.}
\label{fig:pafp}
\end{figure*}

Because the LOS magnetogram around the prominence on May 25, 2012,
close to the west solar limb, shows projection effect and large errors,
we looked back in time of the previous four days when the prominence was closer to the disk
center and used H$\alpha$ images from GONG to locate the prominence legs. We label their
footpoints with small circles in Figure~\ref{fig:pafp}. For each day, we plot the AIA 
171 \AA~images and the HMI LOS magnetograms in the same field of view 
at the closest time of the H$\alpha$ images and plot circles with coordinates found
in the H$\alpha$ footpoints. Circles with the same color in the three panels of a row 
have the same coordinates. The footpoints are labeled G1, G2, and so on from 
west to east from the top to the bottom row with consecutive numbering because we do not
mean to follow the time evolution of prominence legs with this low time cadence.
The 171 \AA~images are in log scale to highlight dark structures. 
The filament legs in H$\alpha$ images correspond well with the filament legs found in 171 \AA~images.
We applied a Gaussian smooth filter on the HMI magnetograms to reduce noise in weak-field regions 
and highlight strong magnetic elements. We outline the supergranular boundaries presented by
magnetic network near the polarity-inversion line (PIL) with dashed lines connecting strong magnetic elements in 
the HMI magnetograms. In the first row on May 20, four out of five footpoints except for G3 are at 
supergranular boundaries. In the second row on May 21, five out of six footpoints except for G8 are at
supergranular boundaries. In the third row on May 22, all seven footpoints are at
supergranular boundaries. In the last row on May 23, nine footpoints except for G19, which
is close to the solar limb with large errors of magetic field measurement, are at 
supergranular boundaries. In this case, we find that about 93\% (25 out of 27) of the recognizable
footpoints are located at supergranular boundaries around the PILs.
There are separations of about one to three supergranular cells between two neighboring footpoints.

\begin{figure*}
\includegraphics[width=\textwidth]{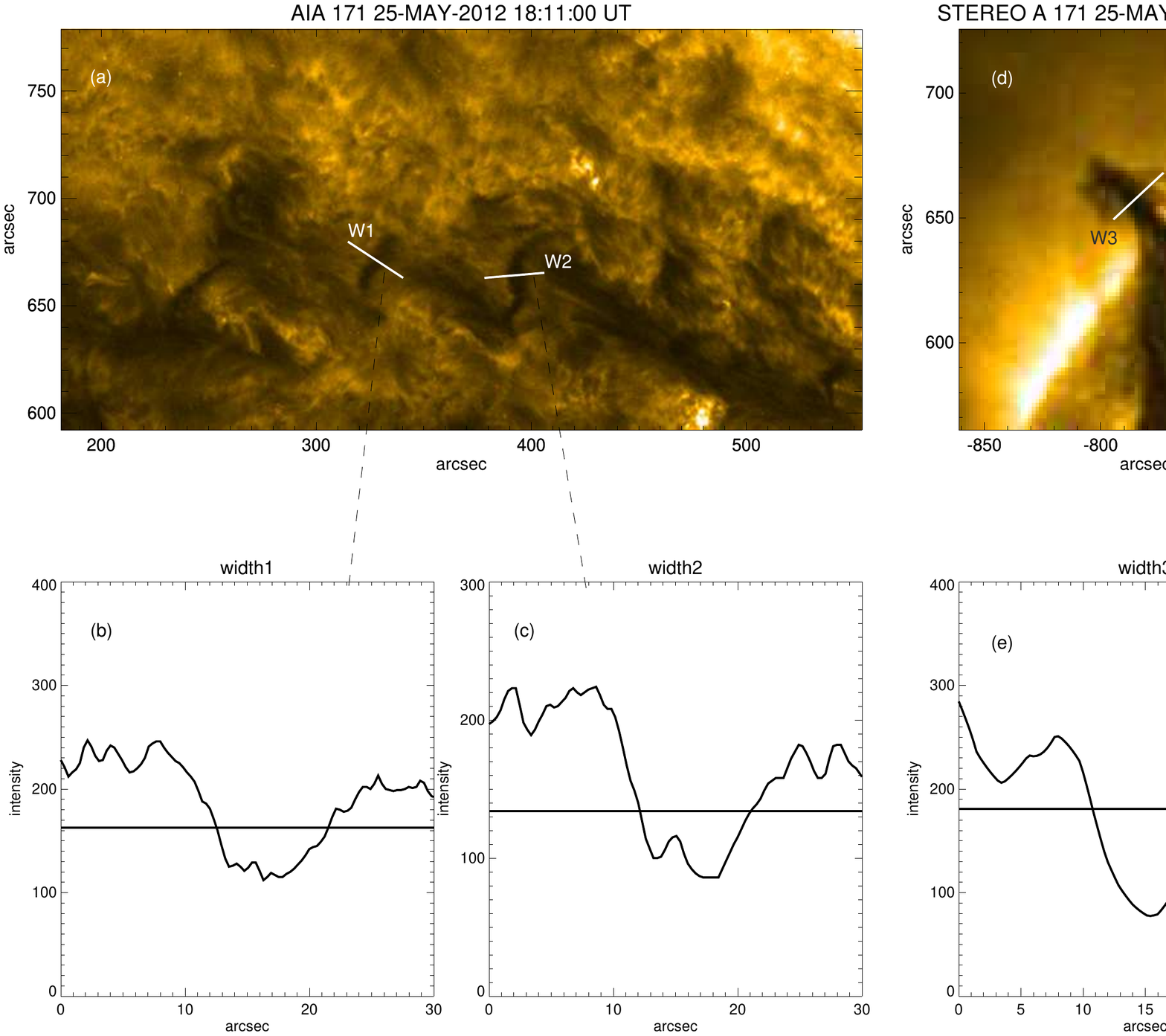}
\caption{Measurement of the width of two legs of prominence PA. (a) two prominence 
legs in the SDO/AIA 171 \AA~image and two cutting lines in white. (b) Intensity curve along
the left cutting line in (a). (c) Intensity curve along the right cutting line in (a).
(d) Two overlapping prominence legs in STEREO A 171 \AA~image with a line cut in 
white. (e) Intensity curve along the line cut in (d). The horizontal solid lines
in (b), (c), and (e) show the intensity value of the leg boundaries and indicate
the widths of the legs with the intersection points.}
\label{fig:paw}
\end{figure*}

To investigate the shape of the cross-sections 
of prominence legs, we measured the apparent width of the PA legs F3-T3 and F4-T4 in two 
viewing angles at the same time. The LOS of SDO, in the right panel of 
Figure~\ref{fig:pa}, is roughly perpendicular to the neutral line, while the LOS of STEREO A 
is along the neutral line. As shown in Figure~\ref{fig:paw}, we drew two lines
that cut perpendicularly through the middle of two legs and plot the intensity curves along the white 
cutting lines. We applied the following procedure to determine the width
of the legs. First, we evaluated the maximum and minimum from the left half 
 and the right half of the intensity curve separately. Second, 
we recorded the smaller of the two maxima and the smaller of the two 
minima. Third, we calculated the arithmetic mean of these two values and plot it as
a horizontal line in the intensity curve plot. The two intersection points
between the horizontal line and the intensity curve are treated as the boundaries 
of the prominence leg. The distance between two boundaries along the cutting line 
is counted as the apparent width of the leg. This measurement of the width is also 
known as the full width at half minimum (FWHM), which was used to measure the width 
of filament threads \citep{Lin2005}. The apparent widths of the two legs in the AIA 
image are of the same value of 8.6$\arcsec$ (6,230 km). The apparent width 
of the two overlapped legs in the STEREO A image is 8.9$\arcsec$ (6,450 km), which is
the upper limit because of the overlapping in the LOS. The width of the legs that appeared
in the LOS along the neutral line is comparable to or smaller than the width 
that appeared in the LOS perpendicular to the neutral line.

\begin{figure*}
\includegraphics[width=\textwidth]{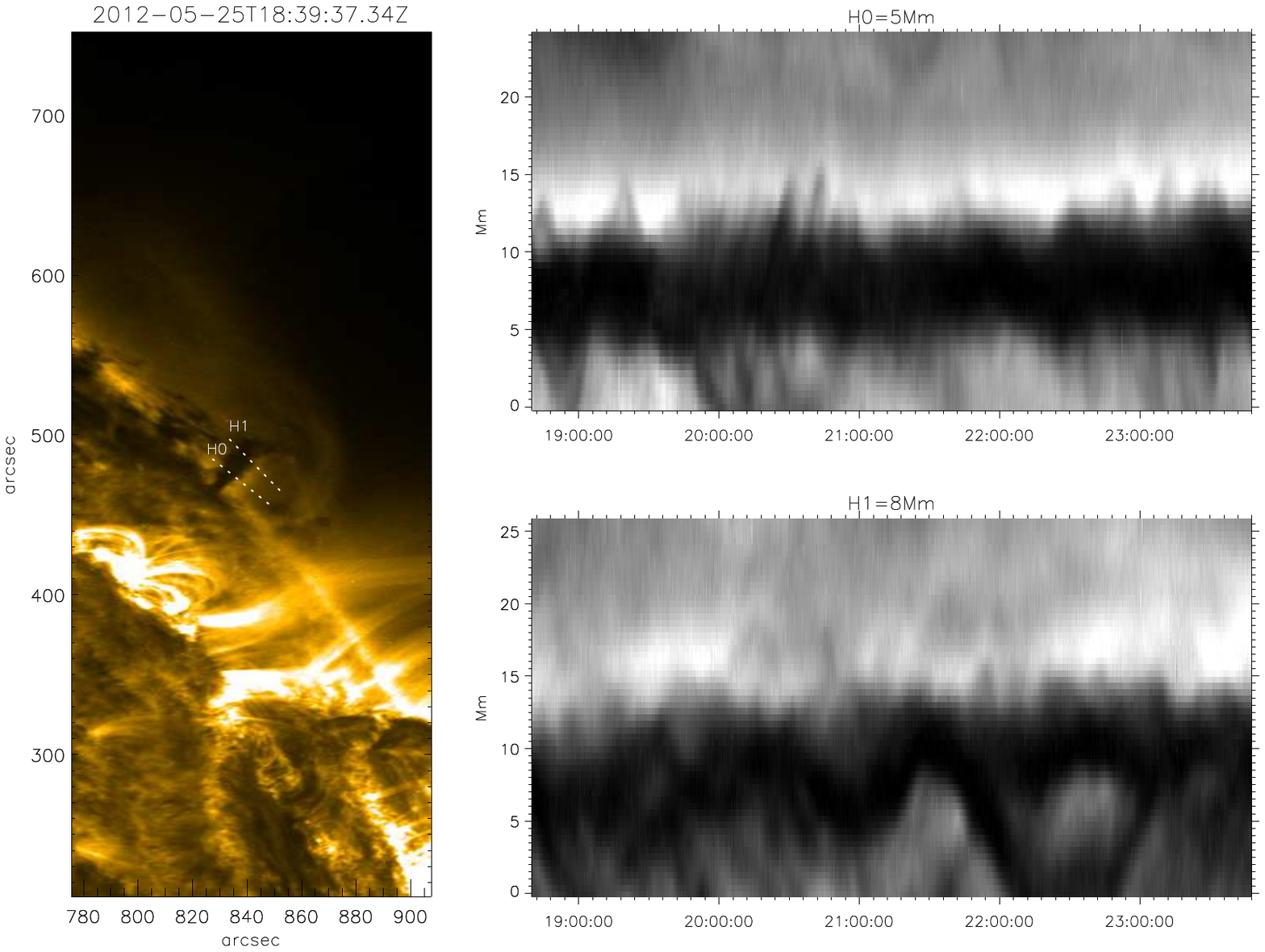}
\caption{Transverse oscillations in a leg of prominence PA. The left panel presents 
the AIA 171 image of partial spine and legs of PA 
two dotted horizontal lines H0 and H1 cutting across a prominence leg above the
solar limb. The right panels are time-distance maps for cuts H0 and H1; the 12-second cadence covers about five hours on May 25, 2012.
}
\label{fig:PAslice}
\end{figure*}

To investigate the apparent motion of fine structures in prominence legs, we drew two fixed cutting 
lines H0 and H1 across the leg F1-T1 above the solar limb in AIA 171 \AA~images and 
plot the time-distance maps of the cuts in Figure~\ref{fig:PAslice}. Both maps show
apparent oscillations of the leg in the horizontal direction nearly perpendicular to
the leg axis. The oscillations in the higher part of the leg show larger amplitudes 
than the oscillations in the lower part, which was also found in a shock-driven 
case of prominence transverse oscillations in previous work \citep{Shen2014}.
This result indicates that the upper part of prominence legs may have shallower
 magnetic dips than the lower part, which agrees with the 3D linear force-free magnetic
field model of \citet{Gunar2018}, assuming that the prominence plasma is moving along
magnetic field lines, 

\subsection{Intermediate-type prominence: PB}\label{sec:PB}

\begin{figure*}
\includegraphics[width=\textwidth]{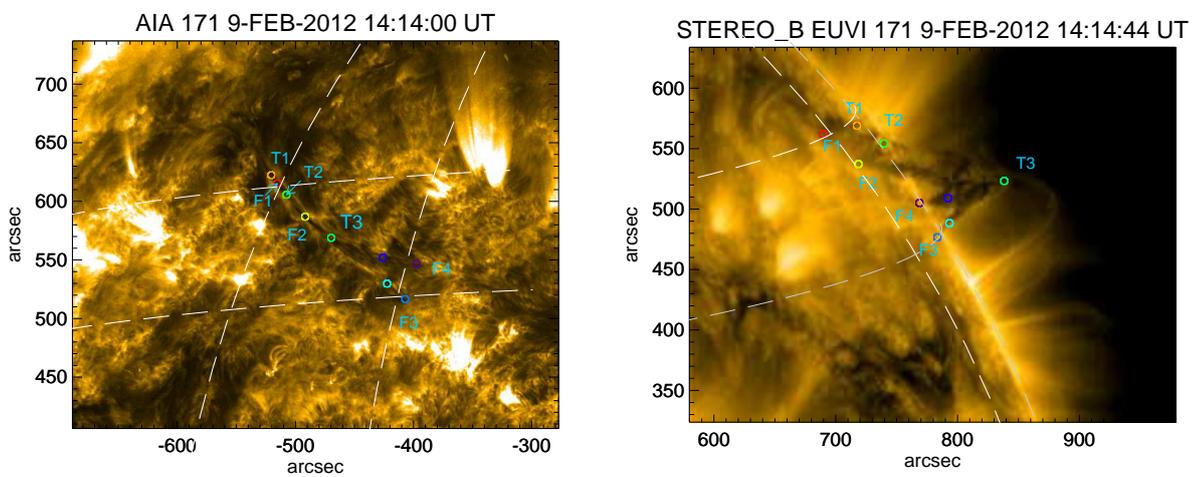}
\caption{Prominence PB observed on February 9, 2012, by SDO/AIA (left) 
and STEREO B/EUVI (right) in 171 \AA~channel. {The EUVI image is in log-scale
for clarity. Different features are marked with circles in different colors, 
and circles of the same color in two images represent the same feature.
The diameter of these circles is 6 arcsec (about 4,350 km).}
Dashed white lines are the Carrington latitude and longitude lines according to 
the latitudes and longitudes of points F1 and F3.
}
\label{fig:PB1}
\end{figure*}

In addition to quiescent prominences, we also investigated the 3D structure of 
an intermediate-type prominence PB that was observed on February 9, 2012,
when the STEREO B and SDO were separated with a viewing angle difference
of 115.85 degrees. As shown in Figure~\ref{fig:PB1}, the prominence was located
in the east part of the solar disk in the SDO/AIA 171 \AA~image, 
while it appeared at the west solar limb in the STEREO B/EUVI 171 \AA~ 
image. The prominence presents a continuous spine structure in the AIA image 
with the LOS looking down. However, from the STEREO B/EUVI LOS
looking from the side, the prominence shows multiple discontinuous internal 
structures, in which we mark four legs as F1-T1, F2-T2, F3-T3, and 
F4-T3. The prominence forks to the leg F3-T3 and the leg F4-T3 at the T3 point,
which is the highest point of the whole prominence.
Two blue circles mark the middle parts of leg F3-T3 and leg F4-T3, respectively.
Reconstructed true sizes of the prominence can be found in Table~\ref{table:1}.
The leg F2-T2 is almost vertical to the solar surface (86 degrees), while 
the legs F1-T1, F3-T3, and F4-T3 have smaller inclination angles relative to the
solar surface, around 60 degrees. The highest point T3 has a height of 
60,181$\pm$1,051 km, and the total length of the prominence is about 297,000 km.

\begin{figure*}
\includegraphics[width=\textwidth]{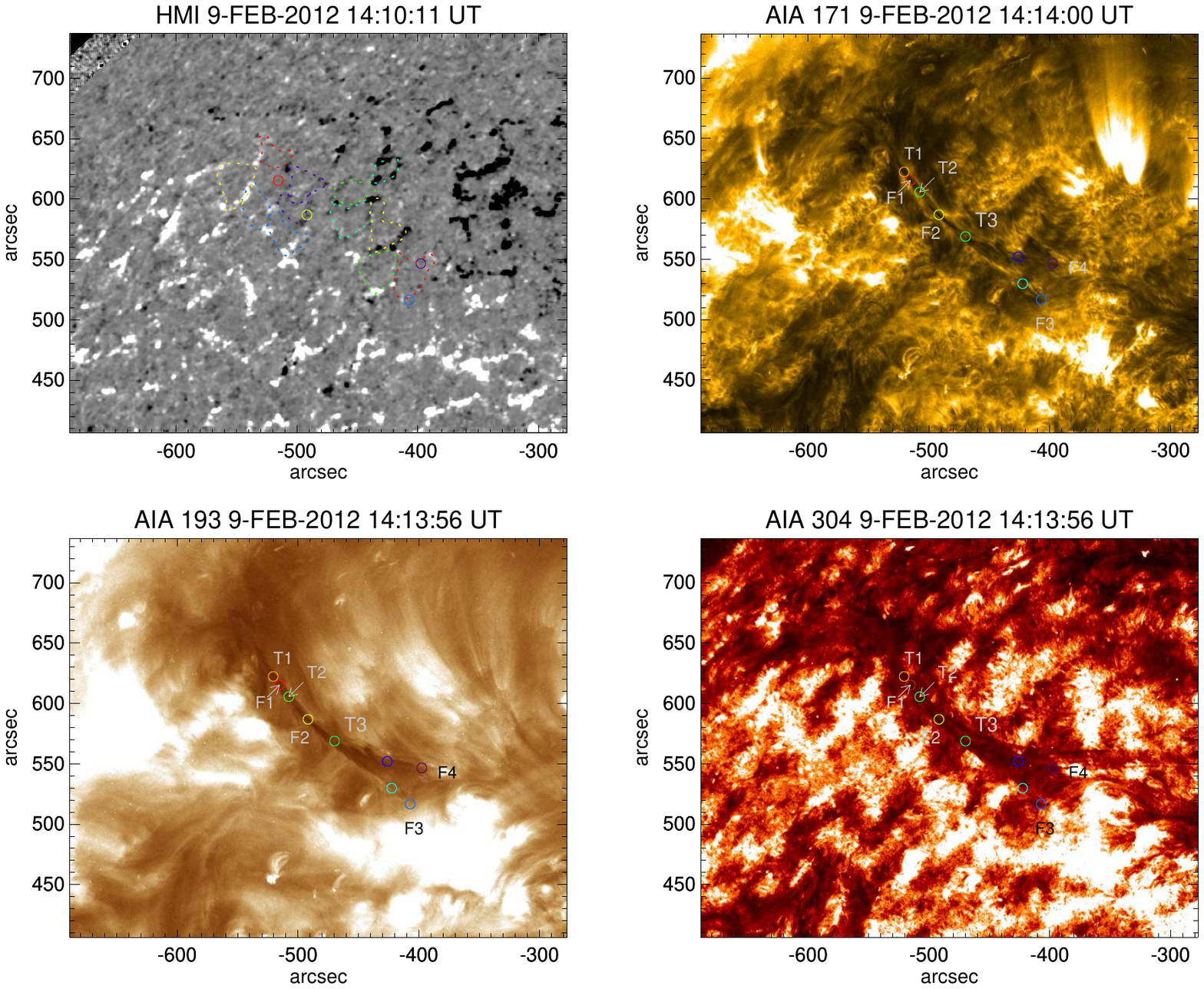}
 \caption{HMI magnetogram, AIA 171 \AA, 193 \AA, and 304 \AA~ images with the same
field of view looking at prominence PB on February 9, 2012. Prominence features
are marked with circles in different colors for all AIA images. Only footpoints
of the prominence are marked with circles in the magnetogram, which is shown in 
gray saturated at $\pm30$ G. Circles of the same color in these images represent 
the same feature. The diameter of these circles is 8 arcsec (about 5,800 km).}
\label{fig:PB2}
\end{figure*}

To investigate the magnetic environment of prominence PB, we plot the HMI 
magnetogram around it, together with AIA 171 \AA, 193 \AA, and 304 \AA~
images in the same field of view in Figure~\ref{fig:PB2}. The magnetogram 
is Gaussian smoothed and saturated at $\pm30$ G to clearly show the 
magnetic network indicated by colored dashed curves near the PIL.
All four reconstructed footpoints F1, F2, F3, and F4 are located at supergranular boundaries 
shown as magnetic network in the polarity inversion region, where network magnetic fluxes with opposite 
polarities mix and neutralize. In the 171 \AA~image, a continuous dark
filament is outlined by bright emission edges against a broad dark lane 
sandwiching the filament. These bright edges may be caused by the emission 
of Fe IX 171.07 \AA~line from the PCTR around 400,000 K. But, there are no 
such bright edges around the filament in the 193 \AA~image, since the PCTR 
temperature is far away from 1.6 MK, which is the dominant line formation 
temperature for the 193 \AA~channel \citep{Odwyer2010}. The dark filament in the
304 \AA~channel apparently has a larger lateral extension compared to the filament in 
171 \AA~channel, especially around point T3.

\begin{figure*}
\includegraphics[width=\textwidth]{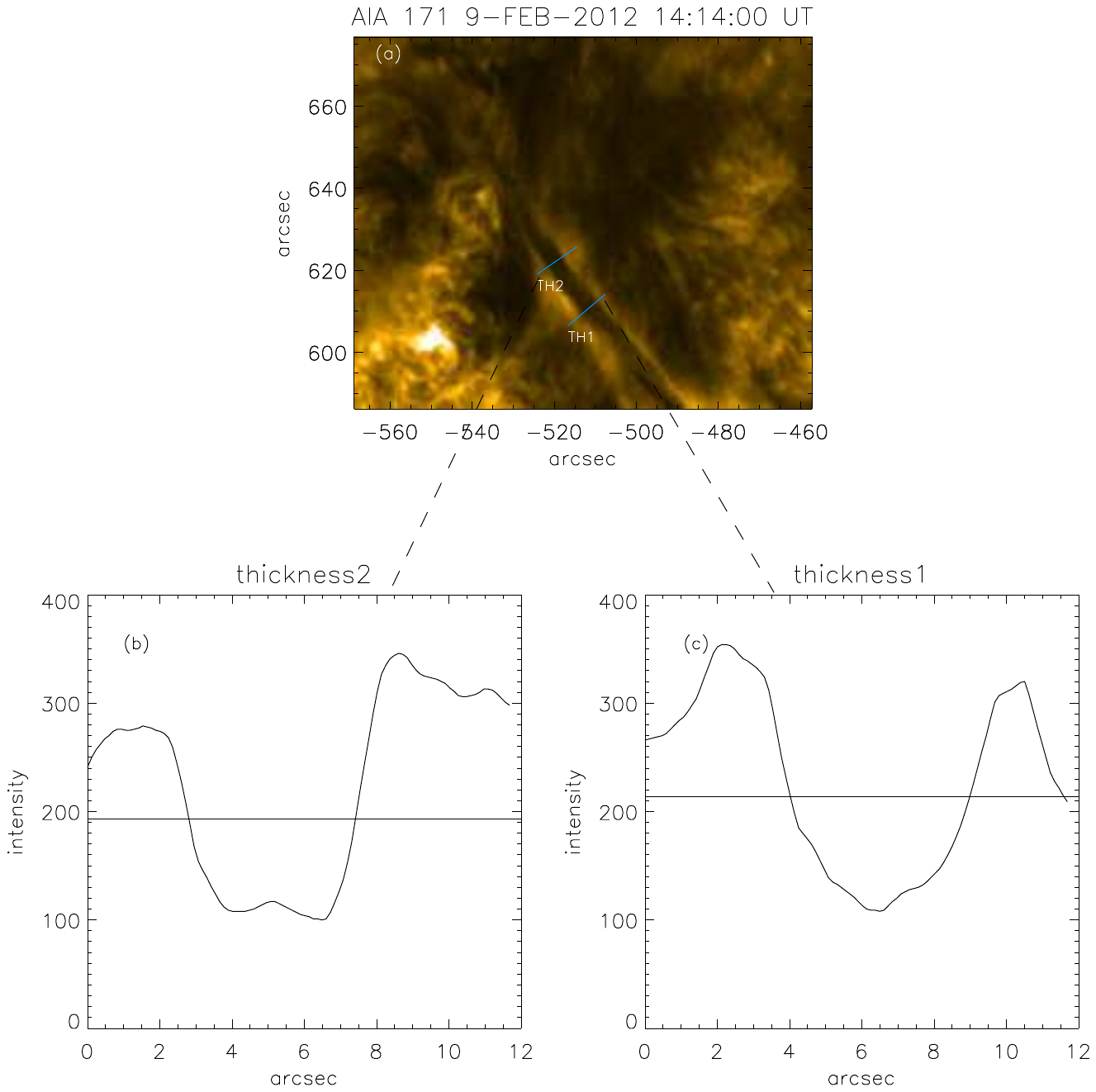}
\caption{Measuring the thickness of the spine of prominence PB. (a) Prominence
spine in the SDO/AIA 171 \AA~image and two lines in blue. (b) Intensity curve along
the upper line in (a). (c) Intensity curve along the lower line in (a).}
\label{fig:PBwidth}
\end{figure*}

Using the FWHM measurement described in Section~\ref{sec:pa}, we obtained 
intensity curves along two lines cutting across the filament spine in the AIA 171 
\AA~image in Figure~\ref{fig:PBwidth}. The lower cut TH1 gives a thickness of 5$\arcsec$ (3,625 km),
and the upper cut TH2 gives a thickness of 4.8$\arcsec$ (3,480 km). The rest of the spine, 
except for the southern bifurcated part, shows a similar thickness as viewed by 
eye. We repeated the width measurement of the filament spine in AIA 193 \AA~image and
AIA 304 \AA~image. The TH1 cut gives 6.5$\arcsec$ and the TH2 cut gives 6.7$\arcsec$ in
the 193 \AA~image, which is about 35\% larger on average than the cuts in 171 \AA. 
The TH1 cut gives 6.4$\arcsec$ and the TH2 cut gives 6.3$\arcsec$ in the 304 \AA~images.
The light curves of the cuts in 304 \AA~image show both emission and absorbtion in 
the filament spine, and the width measurement has large uncertainty caused by the dark 
chromospheric structures around the filament spine in the 304 \AA~channel.
The overall structure of the prominence, except for the southern bifurcated part, 
is reminiscent of in a vertical slab with detailed arch-like structures.

\subsection{Compound-type prominence: PC}\label{sec:PC}

\begin{figure*}
\includegraphics[width=\textwidth]{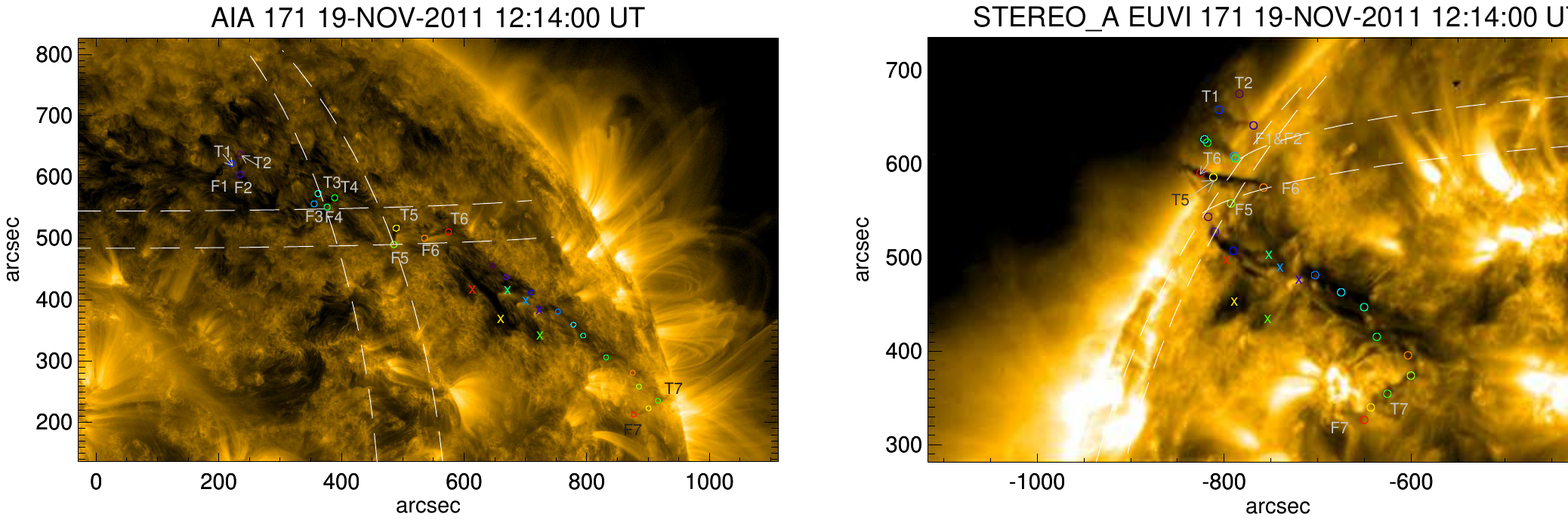}
\caption{Prominence PC observed on November 19, 2011, by SDO/AIA (left) and 
STEREO A/EUVI (right) in the 171 \AA~channel. The EUVI image is in log-scale
for clarity. Different features are 
marked with circles in different colors, and two circles of the same color 
in the two images represent the same feature. 
The diameter of these circles is 8 arcsec (about 5,800 km).
The dashed white lines are the longitude and latitude lines of the Carrington 
coordinates of F4 and F5. The crosses mark some dark patches
and small bright spots in the chromosphere near the prominence
 to avoid confusion.}
\label{fig:PC1}
\end{figure*}

In the third case, we present a 3D reconstruction of the mixed-type 
prominence PC, which was observed on November 19, 2011, when STEREO A
 and SDO were separated by 106.1 degrees. As shown in Figure~\ref{fig:PC1},
 the prominence has a straight distribution extending from the northeast 
quiescent region to the southwest active region, which is characterized by multiple 
coronal loop systems. The northeast part of the prominence is composed of 
six legs, like typical quiescent prominences, while the southwest part is 
dominated by a quasi-continuous spine similar to intermediate-type prominences.
One leg is labeled F7-T7 at the southwest end. The leg is apparently 
vertical in the AIA image, but actually very oblique with a small inclination 
angle of 19$\pm$2 degrees, which may be caused by a strong magnetic field of 
the coronal loops to its east. The legs in the northeast part with a weaker 
magnetic field environment are more vertical with an inclination angle in the 
range of 60-80 degrees, as listed in Table~\ref{table:1}. The highest point 
of the prominence is T6 with an altitude of 45,468$\pm$821 km.
The overall length of the prominence is about 736,000 km.
There are several dark patches and bright features in the chromosphere near 
the prominence, so we mark them with crosses to distinguish them from 
prominence dark features.

\begin{figure*}
\includegraphics[width=\textwidth]{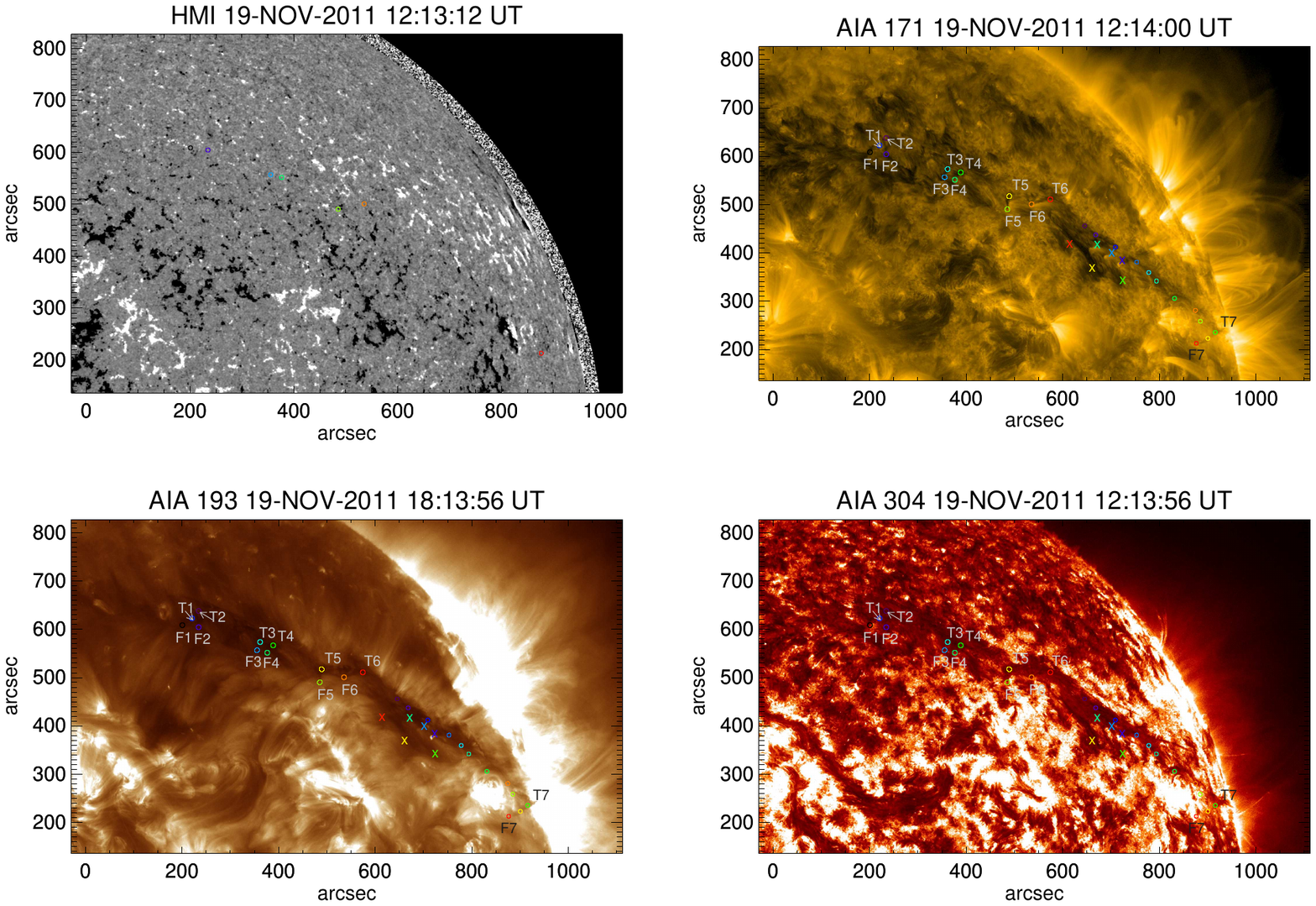}
 \caption{HMI magnetogram, AIA 171 \AA, 193 \AA, and 304 \AA~ images with the same
field of view looking at prominence PC on November 19, 2011. Prominence features 
are marked with circles in different colors for all AIA images. Only footpoints
of the prominence are marked in the magnetogram, which is shown in gray saturated at $\pm40$ G.
The diameter of these circles is 8 arcsec (about 5,800 km).
The crosses mark some dark patches and small bright spots in the chromosphere near the 
prominence to avoid confusion}
\label{fig:PC3}
\end{figure*}

We also plot the HMI photospheric magnetogram around PC with 
the AIA 171 \AA, 193 \AA, and 304 \AA~images in the same field of view 
in Fig~\ref{fig:PC3}. Within seven reconstructed footpoints, six footpoints 
from F1 to F6 are clearly located at supergranular boundaries and the 
footpoint F7 is very close to the limb, with a large uncertainty in the
magnetogram, and it shows a large projection effect. Similar to PA, 
the filament legs are hardly visible in the 193 \AA~image, with 
very poor contrast because the strong foreground emission accumulated 
through the coronal loops, except for leg F7-T7 because its background is bright.
Above the dark chromopheric patches marked with crosses, the filament
spine labeled with purple to blue circles consists of horizontal threads
in the 171 \AA~channel. In the 304 \AA\  channel, the dark threads have lower contrast
and extend above the top of the 171 \AA~filament spine because of the high optical 
thickness, as discussed in Section~\ref{sec:pa}.

\begin{figure*}
\includegraphics[width=\textwidth]{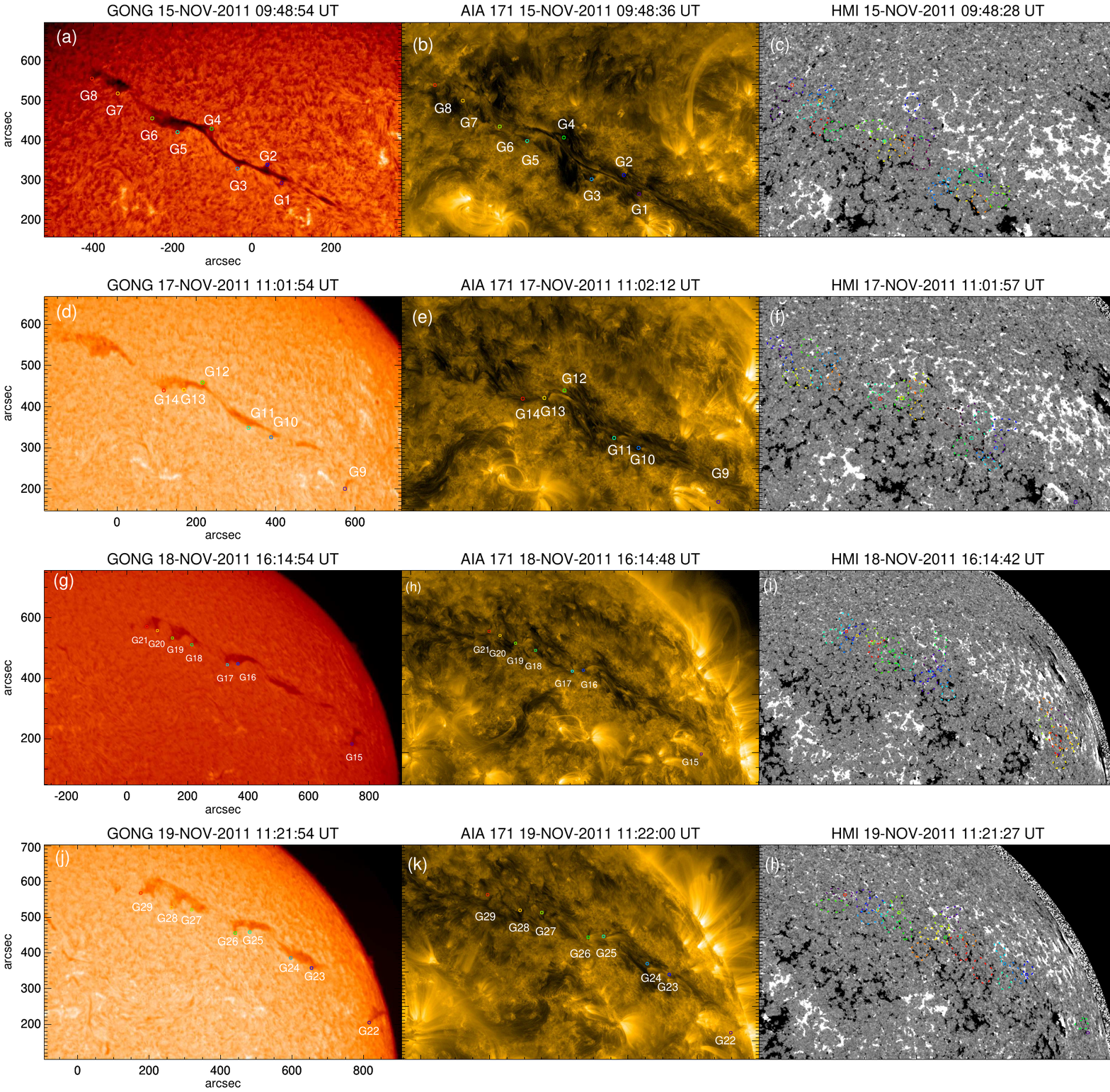}
\caption{Time sequence of prominence PC in H$\alpha$ images, AIA 171 \AA~images in logarithmic scale, 
and LOS magnetograms on November 15 ((a)-(c)), November 17 ((d)-(f)), November 18 ((g)-(i)), and November 19 ((j)-(l))
in 2011. The small circles in different colors show the locations of the footpoints of prominence legs.
The diameter of these circles is 8 arcsec (about 5,800 km).
The magnetograms in gray are Gaussian smoothed and saturated at $\pm30$ G. Supergranular 
boundaries are indicated by dashed lines in different colors connecting strong magnetic 
elements in the HMI magnetograms.}
\label{fig:PC2}
\end{figure*}

To determine the relation between the prominence legs and the underlying magnetic 
features, we again looked back in time to the previous four days and plot the H$\alpha$ images, AIA 171 \AA~images, and LOS magnetograms
side by side, focusing on prominence PC in Figure~\ref{fig:PC2}. The 171 \AA~images are in 
logarithmic scale to highlight dark structures.
A Gaussian smooth filter is applied on the HMI magnetograms to reduce noise and
highlight strong magnetic elements. The filament legs in H$\alpha$ images correspond well with the filament legs found in 171 \AA~images.
We again outline the supergranular boundaries
with dashed lines connecting strong magnetic elements on the HMI magnetograms. 
We mark the tips of barbs and the southern end of legs as footpoints with small circles 
in different colors and ordered labels, as we did for prominence PA. After carefully checking
the circles on the magnetograms, we find that all 29 footpoints in the four snapshots
are at supergranluar boundaries, that is, eight footpoints in the first row on November 15,
six footpoints in the second row on November 17, seven footpoints in the third row on November 18, 
and eight footpoints in the last row on November 19.

\section{Discussion and conclusions}\label{sec:con}

We have used a 3D reconstruction technique on image data of 
simultaneous observations from different viewing angles to study three stable 
prominences: a quiescent, an intermediate, and a mixed type. 
In this section we present our conclusions on the main findings followed 
by further discussion.

We obtained 3D coordinates of the footpoints and top points of prominence 
legs of the three prominences in their stable phase. From these coordinates, 
we derived the inclination angles of the legs relative to the solar surface for the first time. 
These legs are highly inclined and not perpendicular
to the solar surface, although some of them apparently stand radially at 
some viewing angles. The average inclination angle of the 15 legs, except for
the biased PC F7-T7 leg, of the three prominences is 68$\pm$6 degrees.
When the legs from intermediate prominence PB and leg F7-T7 of PC are excluded, 
the average inclination angle of the 11 legs of the quiescent prominence is 68.7$\pm$6 degrees.
Depending on the relative angle with the vertical plane of spines, these inclined legs may
appear to be laterally protruding barbs if the prominences were viewed 
from the top. This picture is consistent with the 3D force-free prominence model 
of \citet{Gunar2018}, in which two laterally protruding filament barbs appear as two inclined 
prominence legs from the synthetic H$\alpha$ prominence view with a LOS perpendicular to the 
prominence axis. The accuracy of our 3D reconstruction is limited by the spatial resolution
of the STEREO EUVI images, and 3D information about fine structures such as filament threads 
is beyond the scope of this study because the resolutions of the AIA and EUVI images are limited.

From the LOS both along and perpendicular to the prominence axis, the apparent widths of the two legs of 
PA have similar values of about 6,000 km, which are consistent with statistic results found 
by \citet{Wedemeyer2013}. The cross-section of the legs is therefore more likely to have a roundish 
or semicircle shape than a thin-sheet shape. The overlapping of the two
legs in the LOS along the filament axis may cause an overestimation of the width. 
Another leg of PA at the solar limb shows apparent horizontal oscillations
with larger amplitude at the higher location, indicating that the upper 
part of prominence legs may have shallower magnetic dips than the lower part.
The cross-section of the legs was studied in only two cases with 
drawbacks from overlap, and  the horizontal oscillations were studied in a single 
case. Therefore a broader study is needed to verify the statistical significance of 
these findings.

By comparing the heights of the top points, we find that different locations 
along a prominence axis have different heights with a two- to threefold
difference, which indicates that the hosting magnetic flux rope may have
fluctuating heights along its axis. The length of the intermediate prominence PB 
(about 290,000 km) is shorter than the quiescent prominence PA (about 560,000 km) and 
the compound quiescent prominence PC (about 730,000 km). Because the footpoints may not be immediately 
beneath a prominence axis, the accumulated distances between footpoints may 
overestimate the true length of the prominence.

To justify the choice of 171\AA~channel images to measure the real filament sizes, we 
present a visual comparison of filament sizes in different spectral bands, for example, AIA 171\AA, 193\AA, 
304\AA~in Figure~\ref{fig:pam} and Figure~\ref{fig:PB2}, and H$\alpha$, AIA 171\AA~in 
Figure~\ref{fig:pafp} and Figure~\ref{fig:PC2}. The sizes of 171\AA~filaments
are similar to the sizes of H$\alpha$ filaments, and the overall structures appear to be
sharper in AIA 171\AA~images given that the AIA 171\AA~images have a better resolution 
than the H$\alpha$ images. Thin filament structures shown in H$\alpha$ images may not be 
visible in the 171\AA~channel with poor contrast due to little absorption of background 
light. No simultaneous observations in H$\alpha$ or other
 chromospheric spectral lines from multiple view angles are available, therefore the best choice for us is to use 
the absorption features in 171\AA~images to represent filaments.
The AIA 304\AA~channel collecting optically thick emission from plasma at about 50,000 K shows
a layer of PCTR in front of prominences, therefore prominence images in 304\AA~channel
above the solar limb occupy a larger volume than the actual prominence plasma does, which was also
 shown by \citet{Berger2011} in their Figure 1.

By checking the footpoint positions in the magnetograms, we find that almost all of the footpoints
in a sample of 70 of the three prominences are located at or very close to the supergranular
boundaries that are traced by magnetic network. This result confirms the previous finding by
\citet{Plocieniak1973}, who suggested that the cool prominence 
plasma in prominence legs is preferentially located at some condensed 
magnetic structure associated with supergranular boundaries. 
Based on 3D reconstruction, we find that all 14 footpoints (excluding the 2 footpoints 
near the solar limb) of the three prominences are at supergranular boundaries. 
To quantify the uncertainty caused by the projection 
effect due to the hight difference between the footpoints and the magnetograms, 
we calculated the projected distance (the distance between the apparent position 
and the radially projected point) of the footpoints in the magnetograms, as listed 
in Table~\ref{table:1}, using the heights and heliocentric angles of the footpoints. 
If we were to consider the projection effect, the footpoints would shift by less than a marking circle 
size (about 5,800 km) toward the solar center because projected distances are comparable to or smaller than the 
diameter of the marking circles. This means that the projection effect does not change our 
conclusion that the footpoints of the three prominences are located near supergranular boundaries. 
About 96\% (54 out of 56) footpoints of the PA and PC based GONG H$\alpha$ images are also found to be 
at supergranular boundaries. The result based on 3D reconstructed footpoints is more 
reliable than the result based on H$\alpha$ images because footpoints found in 
single viewing angle H$\alpha$ images cannot exclude uncertainties from 
projection effects. In order to understand
the magnetic structure of quiescent filament legs, a further dedicated study based on 
high-resolution observations is needed to determine the locations of filament footpoints 
in photospheric magnetograms and investigate possible magnetic flux cancellation near 
filament footpoints. Previous evolutive magnetic flux rope models of quiescent 
prominences \citep{Mackay2006,Xia2014,Xia2016} with a smooth magnetic topology along the PIL need
to consider the effect of supergranulations to further explore possible magnetic 
substructures around prominence legs.

We have measured the inclination angle, the height of top points and footpoints,
 and the length of prominence legs for three prominences. Because the legs of PA and PC are typical 
and representative of legs of quiescent prominences in general, the new knowledge we obtained from 
them is likely to be approximately valid for other quiescent prominences. Although the legs of the intermediate
prominence PB are not so clearly separable as the legs of quiescent prominences, combining the results
from PB can help us to understand general properties for non-active-region stable prominences.
With new space telescopes, such as the Solar Orbiter \citep{Muller2020}
and the Advanced Space-based Solar Observatory (ASO-S) \citep{Gan2019} observing in
different viewing angles in the same waveband, for example, L$\alpha$ images, 
3D reconstruction of prominences will give us more accurate and
detailed 3D information of solar prominences.

\section*{Acknowledgements}
This work is supported by the National Natural Science Foundation of China 
(NSFC, Grant No. 11803031, 11922307, and 11773068), the Basic Research Program of Yunnan 
Province (2019FB140 and 202001AW070011), and Science \& Technology Department of 
Yunnan Province -- Yunnan University Joint Funding (2019FY003005).


\end{document}